\documentclass{elsart}
\usepackage{epsfig}
\usepackage{amsmath}
\usepackage{cite}
\def\Dafne{DA\char8NE}  \def\epm{\ifm{e^+e^-}}
\def\ifm#1{\relax\ifmmode#1\else$#1$\fi}
\def\pic{\ifm{\pi^+\pi^-}}

\def\toP{\ifm{\rightarrow}}
\def\ff{$\phi$ factory}  \def\f{\ifm{\phi}}
\def\Kb{\ifm{\rlap{\kern.2em\raise2ex\hbox to.7em{\hrulefill}} K}}
  \def\ko{\ifm{K^0}}  \def\kob{\ifm{\Kb\kern1pt\vphantom{K}^0}}

\def\ab{\ifm{\sim}}  \def\x{\ifm{\times}}
\def\pt#1,#2,{\ifm{#1\x10^{#2}}}
  
\def\figbox#1;#2;{\parbox{#2\textwidth}{\epsfig{file=#1.eps,width=#2\textwidth}}}
\makeatletter
\newdimen\z@ \z@=0pt 
\newskip\z@skip \z@skip=0pt plus0pt minus0pt
\def\m@th{\mathsurround=\z@}
\def\ialign{\everycr{}\tabskip\z@skip\halign} 
\def\eqalign#1{\null\,\vcenter{\openup\jot\m@th
  \ialign{\strut\hfil$\displaystyle{##}$&$\displaystyle{{}##}$\hfil
    \crcr#1\crcr}}\,}
\makeatother
\newcommand{\aff}[2]{Dipartimento di Fisica dell'Universit\`a #1 e Sezione INFN, #2, Italy.}
\newcommand{\affd}[1]{Dipartimento di Fisica dell'Universit\`a e Sezione INFN, #1, Italy.}
\newcommand{\R}{\ensuremath{R}}
\newcommand{\kl}{\mbox{$K_L$}}
\newcommand{\ks}{\mbox{$K_S$}}
\newcommand{\Pe}{\ensuremath{e}}
\newcommand{\Pem}{\ensuremath{e^-}}
\newcommand{\Pep}{\ensuremath{e^+}}

\newcommand{\Pnu}{\ensuremath{\nu}}
\newcommand{\Pnubar}{\ensuremath{\bar{\nu}}}
\newcommand{\Pphi}{\ensuremath{\phi}}
\newcommand{\Ppi}{\ensuremath{\pi}}

\newcommand{\Ppim}{\ensuremath{\pi^-}}
\newcommand{\Ppip}{\ensuremath{\pi^+}}

\newcommand{\tzero}{\ensuremath{T_{0}}}

\newcommand{\eV}{{e\kern-.07em V}}
\newcommand{\MeV}{{\rm \,M\eV}}
\newcommand{\GeV}{{\rm G\eV}}
\newcommand{\ps}{{\rm \,ps}}
\newcommand{\ns}{{\rm \,ns}}
\newcommand{\mm}{{\rm \,mm}}
\newcommand{\cm}{{\rm \,cm}}
\newcommand{\m}{{\rm \,m}}
\newcommand{\um}{\ensuremath{\mathrm{\mu m}}}
\newcommand{\T}{{\rm \,T}}
\newcommand{\Lpb}{\ensuremath{\rm pb^{-1}}}
\newcommand{\dsdq}{\ensuremath{\Delta S\!=\!\Delta Q}}

\newcommand{\phikskl}{\ensuremath{\phi\rightarrow K_S K_L}}

\newcommand{\DKSeIII}{\ensuremath{K_S\rightarrow\pi e \nu}}

\newcommand{\DKSmuIII}{\ensuremath{K_S\rightarrow\pi \mu \nu}}
\newcommand{\DKSeIIIeppm}{\ensuremath{K_S\rightarrow\Ppim\Pep\Pnu}}

\newcommand{\DKSeIIIempp}{\ensuremath{K_S\rightarrow\Ppip\Pem\Pnubar}}

\newcommand{\DKSpippim}{\ensuremath{K_S\rightarrow\pi^+\pi^-}}
\newcommand{\DKSpippimandg}{\ensuremath{K_S\rightarrow\pi^+\pi^-(\gamma)}}
\newcommand{\DKSpiopio}{\ensuremath{K_S\rightarrow\pi^0\pi^0}}
\newcommand{\DKLeIII}{\ensuremath{K_L\rightarrow\pi e \nu}}

\newcommand{\BR}[1]{\ensuremath{\mathrm{BR}(#1)}}
\newcommand{\gammo}[1]{\ensuremath{\Gamma(#1)}}
\newcommand{\Emiss}{\ensuremath{E_\mathrm{miss}}}
\newcommand{\Pmiss}{\ensuremath{p_\mathrm{miss}}}
\newcommand{\SN}[2]{\ensuremath{#1\times10^{#2}}}
\newcommand{\VS}[2]{\ensuremath{#1\pm#2}}
\newcommand{\Fig}[1]{Fig.~#1}
\newcommand{\Vus}{\ensuremath{V_{us}}}
\newcommand{\Vud}{\ensuremath{V_{ud}}}
\newcommand{\Vub}{\ensuremath{V_{ub}}}

\begin{document}
\begin{frontmatter}
\title{Study of the branching ratio and charge asymmetry for the decay $\DKSeIII$ with the KLOE detector}
\collab{The KLOE Collaboration}
\author[Na]{F.~Ambrosino},
\author[Frascati]{A.~Antonelli}
\author[Frascati]{M.~Antonelli},
\author[Roma3]{C.~Bacci},
\author[Karlsruhe]{P.~Beltrame},
\author[Frascati]{G.~Bencivenni},
\author[Frascati]{S.~Bertolucci},
\author[Roma1]{C.~Bini},
\author[Frascati]{C.~Bloise},
\author[Roma1]{V.~Bocci},
\author[Frascati]{F.~Bossi},
\author[Virginia]{D.~Bowring},
\author[Roma3]{P.~Branchini},
\author[Roma1]{R.~Caloi},
\author[Frascati]{P.~Campana},
\author[Frascati]{G.~Capon},
\author[Na]{T.~Capussela},
\author[Roma3]{F.~Ceradini},
\author[Frascati]{S.~Chi},
\author[Na]{G.~Chiefari},
\author[Frascati]{P.~Ciambrone},
\author[Virginia]{S.~Conetti},
\author[Frascati]{E.~De~Lucia},
\author[Roma1]{A.~De~Santis},
\author[Frascati]{P.~De~Simone},
\author[Roma1]{G.~De~Zorzi},
\author[Frascati]{S.~Dell'Agnello},
\author[Karlsruhe]{A.~Denig},
\author[Roma1]{A.~Di~Domenico},
\author[Na]{C.~Di~Donato},
\author[Pisa]{S.~Di~Falco},
\author[Roma3]{B.~Di~Micco},
\author[Na]{A.~Doria},
\author[Frascati]{M.~Dreucci},
\author[Frascati]{G.~Felici},
\author[Frascati]{A.~Ferrari},
\author[Frascati]{M.~L.~Ferrer},
\author[Frascati]{G.~Finocchiaro},
\author[Roma1]{S.~Fiore},
\author[Frascati]{C.~Forti},
\author[Roma1]{P.~Franzini},
\author[Frascati]{C.~Gatti\corauthref{cor1}},
\author[Roma1]{P.~Gauzzi},
\author[Frascati]{S.~Giovannella},
\author[Lecce]{E.~Gorini},
\author[Roma3]{E.~Graziani},
\author[Pisa]{M.~Incagli},
\author[Karlsruhe]{W.~Kluge},
\author[Moscow]{V.~Kulikov},
\author[Roma1]{F.~Lacava},
\author[Frascati]{G.~Lanfranchi},
\author[Frascati,StonyBrook]{J.~Lee-Franzini},
\author[Karlsruhe]{D.~Leone},
\author[Frascati]{M.~Martini},
\author[Na]{P.~Massarotti},
\author[Frascati]{W.~Mei},
\author[Na]{S.~Meola},
\author[Frascati]{S.~Miscetti},
\author[Frascati]{M.~Moulson},
\author[Frascati,Karlsruhe]{S.~M\"uller},
\author[Frascati]{F.~Murtas},
\author[Na]{M.~Napolitano},
\author[Roma3]{F.~Nguyen},
\author[Frascati]{M.~Palutan},
\author[Roma1]{E.~Pasqualucci},
\author[Roma3]{A.~Passeri},
\author[Frascati,Energ]{V.~Patera},
\author[Na]{F.~Perfetto},
\author[Roma1]{L.~Pontecorvo},
\author[Lecce]{M.~Primavera},
\author[Frascati]{P.~Santangelo},
\author[Roma2]{E.~Santovetti},
\author[Na]{G.~Saracino},
\author[Frascati]{B.~Sciascia},
\author[Frascati,Energ]{A.~Sciubba},
\author[Pisa]{F.~Scuri},
\author[Frascati]{I.~Sfiligoi},
\author[Frascati]{T.~Spadaro\corauthref{cor2}},
\author[Roma1]{M.~Testa},
\author[Roma3]{L.~Tortora},
\author[Frascati]{P.~Valente},
\author[Karlsruhe]{B.~Valeriani},
\author[Frascati]{G.~Venanzoni},
\author[Roma1]{S.~Veneziano},
\author[Lecce]{A.~Ventura},
\author[Frascati]{R.~Versaci},
\author[Frascati,Beijing]{G.~Xu},
\address[Virginia]{Physics Department, University of Virginia, Charlottesville, VA, USA.}
\address[Frascati]{Laboratori Nazionali di Frascati dell'INFN, Frascati, Italy.}
\address[Karlsruhe]{Institut f\"ur Experimentelle Kernphysik, Universit\"at Karlsruhe, Germany.}
\address[Lecce]{\affd{Lecce}}
\address[Na]{Dipartimento di Scienze Fisiche dell'Universit\`a ``Federico II'' e Sezione INFN, Napoli, Italy}
\address[Energ]{Dipartimento di Energetica dell'Universit\`a ``La Sapienza'', Roma, Italy.}
\address[Roma1]{\aff{``La Sapienza''}{Roma}}
\address[Roma2]{\aff{``Tor Vergata''}{Roma}}
\address[Roma3]{\aff{``Roma Tre''}{Roma}}
\address[Pisa]{\affd{Pisa}}
\address[StonyBrook]{Physics Department, State University of New York at Stony Brook, NY, USA.}
\address[Beijing]{Permanent address: Institute of High Energy Physics, CAS, Beijing, China.}
\address[Moscow]{Permanent address: Institute for Theoretical and Experimental Physics, Moscow, Russia.}
\begin{flushleft}
\corauth[cor1]{cor1}{\small $^1$ Corresponding author: Claudio Gatti
INFN - LNF, Casella postale 13, 00044 Frascati (Roma), 
Italy; tel. +39-06-94032727, e-mail claudio.gatti@lnf.infn.it}
\end{flushleft}
\begin{flushleft}
\corauth[cor2]{cor2}{\small $^2$ Corresponding author: Tommaso Spadaro
INFN - LNF, Casella postale 13, 00044 Frascati (Roma), 
Italy; tel. +39-06-94032698, e-mail tommaso.spadaro@lnf.infn.it}
\end{flushleft}
\begin{abstract}
Among some 400 million \ks\kl\ pairs produced in \epm\ annihilations at \Dafne, \ab\,6500 each of \ks\toP$\pi^+ e^-\bar\nu$ and \ks\toP$\pi^- e^+\nu$ decays have been observed with the KLOE detector. From these,
the ratio 
$\gammo{\DKSeIII}/\gammo{\DKSpippim}\!=\!\pt(10.19\pm0.13),-4,$ 
is obtained, improving the accuracy on \BR{\DKSeIII} by a factor of four and 
providing the most precise test of the $\Delta S\!=\!\Delta Q$ rule.
From the partial width \gammo{\DKSeIII}, a value for $f_+^{K^0}(0)\!\x\!\Vus$ is obtained that is 
in agreement with unitarity of the quark-mixing matrix. The lepton charge asymmetry
$A_{S}\!=\!\pt(1.5\pm9.6_{\mathrm{stat}}\pm2.9_{\mathrm{syst}}),-3,$ is compatible with the 
requirements of $CPT$ invariance. The form-factor slope 
agrees with recent results from semileptonic \kl\ and $K^+$ decays. 
These are the first measurements of the charge asymmetry and form-factor slope for semileptonic \ks\ decays.
\end{abstract}
\end{frontmatter}

\section{Introduction}

Semileptonic kaon decays provide at present the best way to learn about $s,\:u$ quark couplings and allow tests of many fundamental 
aspects of the Standard Model (SM). Only the vector part of the weak current has a non-vanishing 
matrix element between a kaon and a pion. The vector current is ``almost'' conserved. For a vector interaction, there are no 
$SU(3)$-breaking corrections to first order in the $s$-$d$ mass difference~\cite{Ademollo:1964sr}, making calculations of hadronic 
matrix elements more reliable. Therefore, the CKM matrix element \Vus\ can be accurately extracted from the 
measurement of the semileptonic decay widths and
the most precise test of unitarity of the CKM matrix can be obtained
from the first-row constraint: $1\!-\!\Delta\!\simeq\!|\Vud|^{2}\!+\!|\Vus|^{2}$.
Using \Vud\ from $0^{+}\!\to\!0^{+}$ nuclear beta decays, a test of the expectation $\Delta\!=\!0$ with a precision of one part per mil
can be performed. 

At a \ff\ very large samples of tagged, monochromatic \ks\ mesons are available. We have
isolated a very pure sample of \ab 13\,000 semileptonic \ks\ decay events and
measured for the first time the partial decay rates for transitions to final states of each charge, \gammo{\ks\!\toP\!\Pep\Ppim\Pnu} and 
\gammo{\ks\!\toP\!\Pem\Ppip\Pnubar}, and the charge asymmetry
\begin{equation}
A_{S}=
\frac{
  \Gamma\left(\ks\toP\Ppim\Pep\nu\right) -
  \Gamma\left(\ks\toP\Ppip\Pem\bar{\nu}\right)
}{
  \Gamma\left(\ks\toP\Ppim\Pep\nu\right) +
  \Gamma\left(\ks\toP\Ppip\Pem\bar{\nu}\right)
}.
\label{eq:asy}
\end{equation}
The comparison of $A_S$ with the asymmetry $A_L$ for \kl\ decays 
allows tests of the $CP$ and $CPT$ symmetries. Comparison of the \ks\ and \kl\ widths $\Gamma(\DKSeIII)$ and $\Gamma(\DKLeIII)$ allows
a test of the validity of the $\Delta S\!=\!\Delta Q$ rule.
Assuming $CPT$ invariance, $A_{S}\!=\!A_{L}\!=\!2\,\mathrm{Re}\,\epsilon\!\simeq\!3\!\times\!10^{-3}$, where $\epsilon$
gives the $CP$ impurity of the \ks, \kl\ mass eigenstates due to $CP$ violation in $K\!\leftrightarrow\!\overline{K}$ $\Delta S=2$ transitions. 
The difference between the charge asymmetries,
\begin{equation}
A_{S}-A_{L}=4\,\left(\mathrm{Re}\,\delta+\mathrm{Re}\,x_{-}\right),
\label{eq:rexm}
\end{equation}
signals $CPT$ violation either in the mass matrix ($\delta$ term), or in the decay amplitudes with $\Delta S\!\neq\!\Delta Q$ ($x_{-}$ term).
The sum of the asymmetries, 
\begin{equation}
A_{S}+A_{L}=4\left(\mathrm{Re}\,\epsilon-\mathrm{Re}\,y\right),
\label{eq:rey}
\end{equation}
is related to $CP$ violation in the mass matrix ($\epsilon$ term) and to $CPT$ violation in the $\Delta S\!=\!\Delta Q$ 
decay amplitude ($y$ term).
Finally, the validity of the $\Delta S\!=\!\Delta Q$ rule in $CPT$-conserving transitions can be tested through the quantity:
\begin{equation}
\mathrm{Re}\,x_{+} = \frac{1}{2}\frac{\Gamma(\DKSeIII)-\Gamma(\DKLeIII)}{\Gamma(\DKSeIII)+\Gamma(\DKLeIII)}.
\label{rex}
\end{equation}

Writing the \ko\ and \kob\ decay amplitudes for final states of each charge as
${\mathcal A}_{\pm}\!=\!A(\ko\!\toP\!e^{\pm}\pi^{\mp}\nu(\Pnubar))$ and
$\bar{{\mathcal A}}_{\pm}\!=\!A(\kob\!\toP\!e^{\pm}\pi^{\mp}\nu(\Pnubar))$,
the above parameters are defined as follows:
\begin{subequations}
\begin{equation}\eqalign{
x_{\pm} & = \frac{1}{2}\left[
\frac{\bar{{\mathcal A}}_{+}}{{\mathcal A}_{+}} \pm \left(\frac{{\mathcal A}_{-}}{\bar{{\mathcal A}}_{-}}\right)^{\ast}
\right], \cr
    y & = \frac
{\bar{{\mathcal A}}_{-}^{\ast}-{\mathcal A}_{+}}
{\bar{{\mathcal A}}_{-}^{\ast}+{\mathcal A}_{+}}, \cr}
\end{equation}
\begin{equation}\eqalign{
 \epsilon & = i\frac{\mathrm{Im}M_{12}-i\mathrm{Im}\Gamma_{12}/2}{m_{S}-m_{L}-i(1/\tau_S -1/\tau_L )/2},\cr
 \delta & = \frac{1}{2}\frac{M_{11}-M_{22}-i(\Gamma_{11}-\Gamma_{22})/2}{m_{S}-m_{L}-i(1/\tau_S -1/\tau_L )/2},\cr}
\label{eq:definitions}
\end{equation}
\end{subequations}
where $M_{ij}$ and $\Gamma_{ij}$ are the 
elements of the mass and decay matrices describing the time evolution of the neutral kaon system,
and $m_{S,L}$ and $\tau_{S,L}$ are respectively the 
masses and lifetimes for $K_{S,L}$.

The value of $A_{L}$ is known at present with an accuracy of
$10^{-4}$~\cite{KTeVasy:02}, while $A_{S}$ has never yet been measured.
At present, the most precise test of $CPT$ conservation
comes from the CPLEAR experiment~\cite{CPLEAR_redelta:98}: they find 
$\mathrm{Re}\,\delta$ and $\mathrm{Re}\,x_{-}$  
to be compatible with zero, with accuracies of $3\!\times\!10^{-4}$ and $10^{-2}$, respectively. 
The present value of $\mathrm{Re}\,y$, obtained from unitarity~\cite{CPLEAR:bell}, 
is compatible with zero 
to within $3\!\times\!10^{-3}$. 

In the SM, $\mathrm{Re}\,x_{+}$ is on the order of $G_{F} m_{\pi}^2\!\sim\!10^{-7}$,
being due to second-order weak transitions. 
At present, the most precise test of the \dsdq\ rule
comes from an analysis of the time distribution of strangeness-tagged semileptonic kaon decays at 
CPLEAR~\cite{CPLEAR_rex:98}. 
They found
$\mathrm{Re}\,x_{+}$ to be compatible with zero to within $6\!\times\!10^{-3}$.

The most precise previous measurement of \BR{\DKSeIII} was 
obtained by KLOE
using $\sim20\!$pb$^{-1}$ of data collected in 2000 and has a fractional accuracy of 5.4\%~\cite{Aloisio:2002rq}.
The present result is  
based on the analysis of 410~pb$^{-1}$ of data and
improves on the total error by a factor of four, to 1.3\%.
\section{Measurement method}
We measure \ks\ branching ratios using kaons from \phikskl\ decays.
The data were collected with the KLOE detector at \Dafne, the Frascati \f-factory. \Dafne\ 
is an \epm\ collider 
that operates 
at a center of mass energy of \ab1020\MeV, the mass of the \f\ meson. 
Equal-energy positron and electron beams collide at an angle of $\pi-25$\,mrad, producing \f\ mesons with a
small momentum in the horizontal plane: $p_{\phi}\!\ab13\!\MeV.$
\f\ mesons decay $\ab\!34\%$ of the time into neutral kaons.
\kl's and \ks's have mean decay paths of $\lambda_{L}\!\ab\!350$\cm\ and
$\lambda_{S}\!\ab\!0.6$\cm, respectively.

The KLOE detector
consists of a large cylindrical drift chamber surrounded by a lead/scintillating-fiber
sampling calorimeter. A superconducting coil outside the calorimeter 
provides a 0.52\T\ field. The drift chamber~\cite{DCnim}, 
which is
4\m\ in diameter and 3.3\m\ long, has 
12\,582 
cells strung in all-stereo geometry.
The chamber shell is 
made of a carbon-fiber/epoxy composite. The chamber is filled with a 90\% He, 10\% iC$_4$H$_{10}$ mixture. 
These features maximize transparency to photons and reduce \kl\toP\ks\ regeneration. 
The spatial resolutions are $\sigma_{xy}\!\sim\!150\,\um$ and $\sigma_z\!\sim\!2\mm.$ The momentum resolution is 
$\sigma_{p_{\perp})}/p_{\perp}\!\leq\!0.4\%$. Vertices are reconstructed with a spatial resolution of $\sim\!3\mm$. 
The calorimeter~\cite{EmCnim} is divided into a barrel and two endcaps and covers 98\% of the 
solid angle. The energy resolution is $\sigma_E/E\!=\!5.7\%/\sqrt{E (\GeV)}$ and the timing resolution 
is $\sigma_t\!=\!57\ps/\sqrt{E (\GeV)}\oplus100\ps$~\cite{pennote2}. 
The trigger used for the present analysis
relies entirely on calorimeter information~\cite{TRGnim}. Two local energy deposits above threshold 
($50$\MeV\ on the barrel, $150$\MeV\ on the endcaps) are required.
The trigger time has a large spread with respect to the bunch-crossing period. However, it is 
synchronized with the machine RF divided by 4, $T_\mathrm{sync}\!\sim\!10.8\ns$, with an accuracy of 50\ps. 
As a result,
the time
\tzero\ of the bunch crossing producing an event, which is determined after event reconstruction, is known up to
an integer multiple of the bunch-crossing time, $T_\mathrm{bunch}\!\sim\!2.7\ns$.

The main advantage of studying kaons at a \ff\ is that \kl's and \ks's are produced nearly
back-to-back in the laboratory so that detection of a \kl\ meson tags the 
production of a \ks\ meson and gives its direction and momentum. 
The contamination from $\kl\!\kl\!\gamma$ and $\ks\!\ks\!\gamma$ final states
is negligible for our measurement~\cite{Paver:1990fn,brownclose}.
Since the branching ratio for \DKSpippim\ is known with an accuracy 
of $\ab0.1\%$~\cite{PDBook,plbrapponew}, the
\DKSeIII\ branching ratio is evaluated by normalizing the number of signal 
events, separately for each charge state, to the number of \DKSpippim\
events in the same data set. 
This allows cancellation of the uncertainties arising
from the integrated
luminosity, the \f\ production cross section, and the tagging efficiency.
The measurement is based on an integrated luminosity of
410\,\Lpb\ at the \f\ peak collected during two distinct data-taking periods in
the years 2001 and 2002, corresponding to 
$\ab\!\pt1.3,9,$ produced \f-mesons. Since the machine conditions were 
different during the two periods, we have measured the branching ratios separately for
each data set. Our final results are based on the averages of these measurements.
\section{Selection criteria}

About half of the \kl\ mesons reach the calorimeter, where most interact. 
Such an interaction is called a \kl\ {\it crash} in the
following. A \kl\ crash is identified as a local energy deposit with $E\!>\!200\MeV$
and a time of flight corresponding to a low velocity: 
$\beta\!\ab\!0.216$. The coordinates of the energy deposit determine the \kl\
direction to within $\ab\!20$\,mrad, as well as the momentum $\mathbf{p}_L$,
which is weakly dependent on the \kl\ direction because of the motion of the \f\ meson. 
A \kl\ crash thus tags the production of a \ks\ of momentum
$\mathbf{p}_S\!=\!\mathbf{p}_\Pphi-\mathbf{p}_L$. 
\ks\ mesons are tagged with an overall efficiency of $\ab\!20\%$.
Both \DKSeIII\ and \DKSpippim\ decays are selected from this tagged sample. 
Event selection consists of fiducial cuts,
particle identification by time of flight, and kinematic closure.

Identification of a \DKSpippim\ decay requires two tracks of opposite curvature. 
The tracks must extrapolate to the interaction
point (IP) to within a few centimeters. The reconstructed momenta and polar angles 
must lie in the intervals 
$120\MeV\!<\!p\!<\!300\MeV$ and $30^{\circ}\!<\!\theta\!<\!150^{\circ}$. 
A cut in $(p_{\perp},p_{\parallel})$ selects 
non-spiralling tracks. The numbers of \DKSpippim\ events found in each data set
are shown in Tab.~\ref{yields}.
Contamination due to \ks\ decays other than \DKSpippim\
is at the per-mil level and is estimated from Monte Carlo (MC).
\begin{table}
    \centering
\renewcommand{\arraystretch}{1.1}    
    \begin{tabular}{ccc} \hline
                                & Year 2001 & Year 2002   \\ 
      Luminosity, \Lpb          & \VS{152}{1}   & \VS{286}{2} \\ \hline
      Channel                   & \multicolumn{2}{c}{Number of selected events} \\ \hline
      \DKSpippim                & 13\,056\,500  & 22\,840\,700 \\ \hline
      $\ks\to\Ppip\Pem\Pnubar$ &  \VS{2387}{\VS{52_{\mathrm{stat}}}{22_{\mathrm{syst}}}}  &  \VS{4238}{\VS{69_{\mathrm{stat}}}{55_{\mathrm{syst}}}} \\ 
      $\ks\to\Ppim\Pep\Pnu$    &  \VS{2541}{\VS{52_{\mathrm{stat}}}{24_{\mathrm{syst}}}}  &  \VS{4446}{\VS{71_{\mathrm{stat}}}{69_{\mathrm{syst}}}} \\ \hline
    \end{tabular}
    \caption{Number of selected \DKSpippim\ and \DKSeIII\ decays for the 2001 and 2002 data sets.}
  \label{yields}
\end{table}

Identification of a \DKSeIII\ event also begins with the requirement of 
two tracks of opposite curvature.
The tracks must extrapolate {\it and form a vertex} close to the IP. 
The invariant mass $M_{\pi\pi}$ of the pair calculated 
assuming both tracks are pions must be smaller 
than 490\MeV. This rejects $\ab\!95\%$ of the \pic\ decays
and $\ab\!10\%$ of the signal events.

\begin{figure}[ht]
 \center
    \epsfig{file=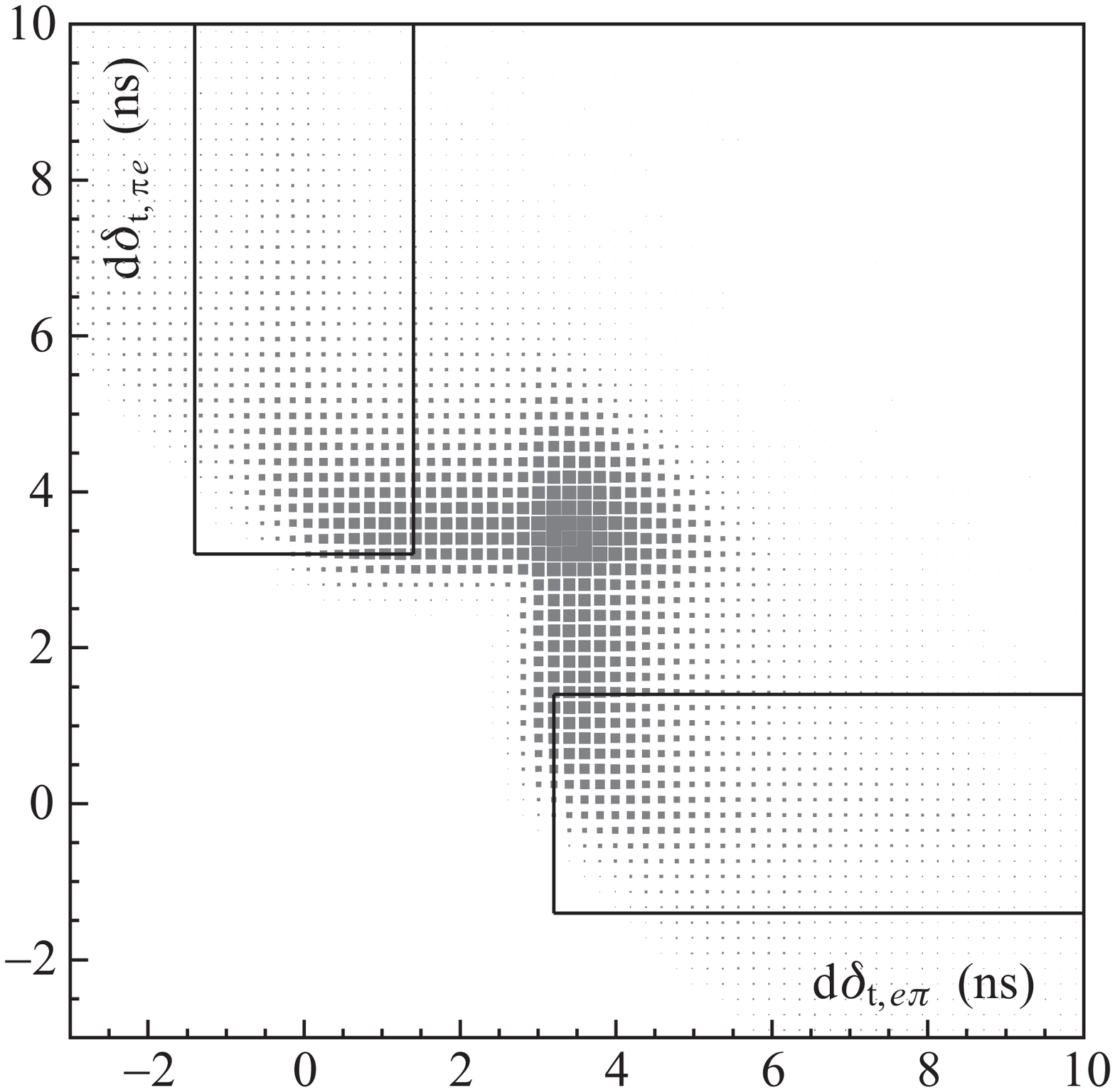, width=0.48\textwidth}
    \epsfig{file=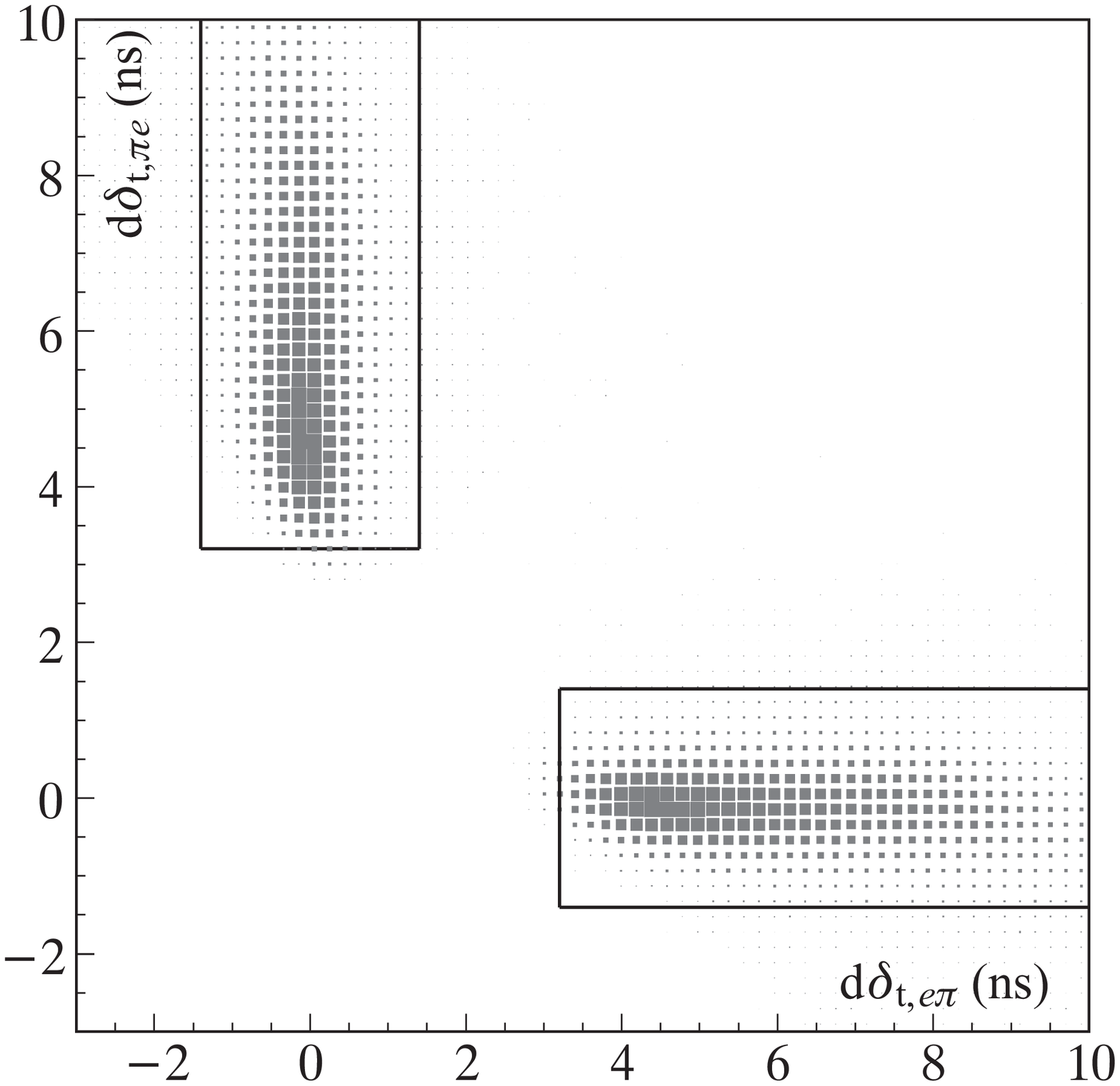, width=0.48\textwidth}
\caption{Scatter plot of the time differences 
$\mathrm{d}\delta_{t,\Ppi\Pe}$ vs $\mathrm{d}\delta_{t,\Pe\Ppi}$
for Monte Carlo events, for all \ks\ decays (left)
and for \DKSeIII\ decays (right).}
\label{cut}
\end{figure}

We discriminate between electron and pion tracks by time of flight (TOF). 
The tracks are therefore required to be
associated with calorimeter energy clusters. For each track,
we compute the difference $\delta_t(m)\!=\!t_\mathrm{cl}-L/c\beta(m)$
using the cluster time $t_\mathrm{cl}$ and the track length $L$. The velocity
$\beta$ is computed from the track momentum for each mass hypothesis, 
$m\!=\!m_{e}$ and $m\!=\!m_{\Ppi}$.
In order to avoid uncertainties due to the determination of \tzero\
(the time of the bunch crossing producing the event),
we make cuts on the two-track difference
$$
\mathrm{d}\delta_{t,ab}=\delta_t(m_{a})_{1}-\delta_t(m_{b})_{2}\mbox{,}
$$
where the mass hypothesis $m_{a(b)}$ is used for track 1(2). 
This difference is zero for the correct mass assignments.
First, \DKSpippim\ events are rejected by requiring
$|\mathrm{d}\delta_{t,\Ppi\Ppi}|\!>\!1.7$\ns. Then, the differences
$\mathrm{d}\delta_{t,\Ppi\Pe}$ and $\mathrm{d}\delta_{t,\Pe\Ppi}$
are calculated for surviving events. The scatter plot
of the two variables is shown in \Fig{\ref{cut}} for Monte Carlo events. The cuts applied on
these time differences for the selection of \DKSeIII\ events are illustrated in the figure:
$|\mathrm{d}\delta_{t,\Ppi\Pe}|\!<\!1.4$\ns, $\mathrm{d}\delta_{t,\Pe\Ppi}\!>\!3.2$\ns; or
$|\mathrm{d}\delta_{t,\Pe\Ppi}|\!<\!1.4$\ns, $\mathrm{d}\delta_{t,\Ppi\Pe}\!>\!3.2$\ns.
After these TOF requirements, particle types and charges for signal events can be assigned
very precisely: the probability of misidentifying a $\Ppip\Pem\Pnubar$ 
event as $\Ppim\Pep\Pnu$ or vice versa is negligible.
These cuts reject $\ab\!90\%$ of the background events, while
the efficiency for the signal is $\ab\!85\%$.

\begin{figure}[ht]
 \center
  \epsfig{file=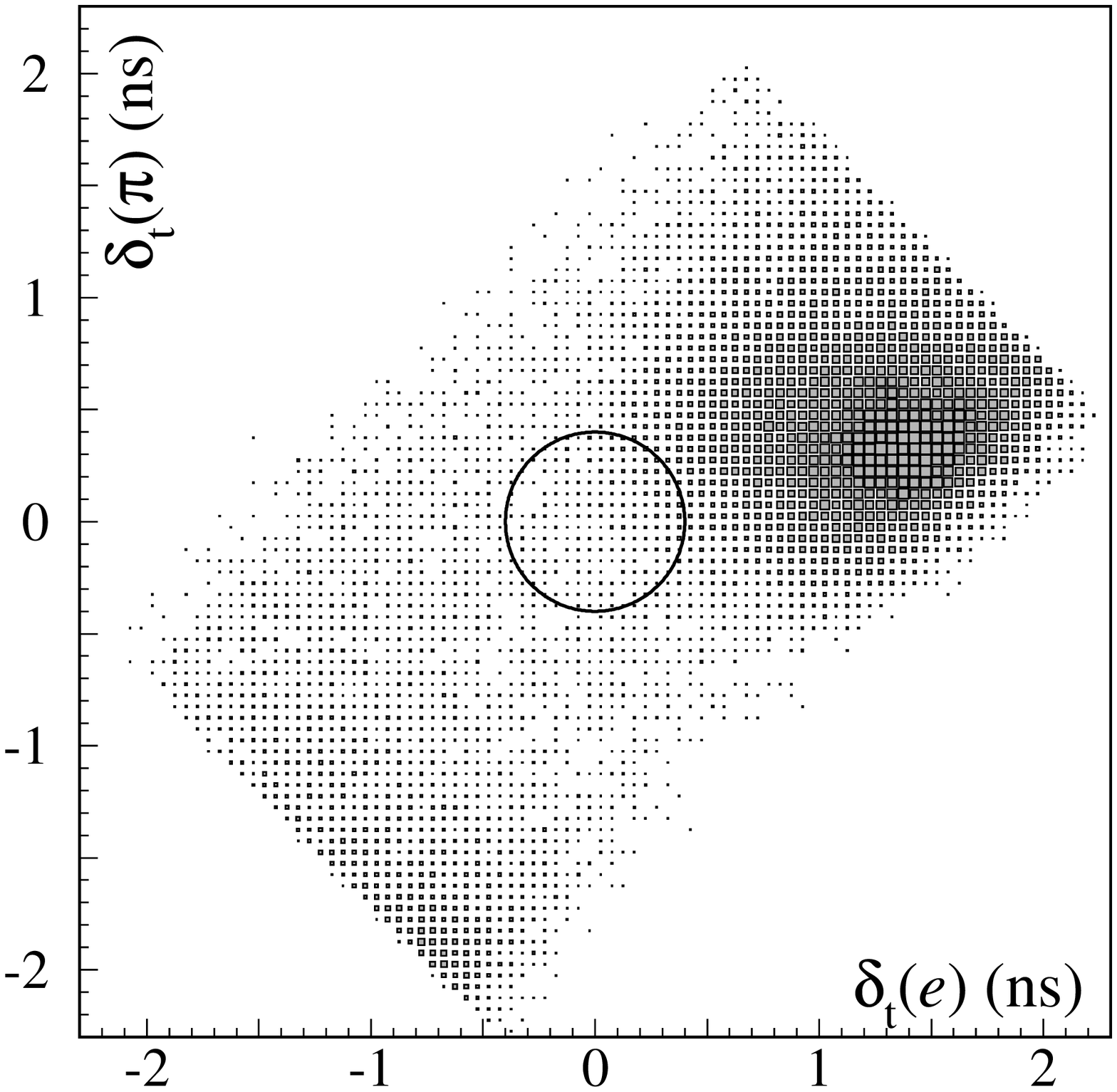, width=0.48\textwidth}
  \epsfig{file=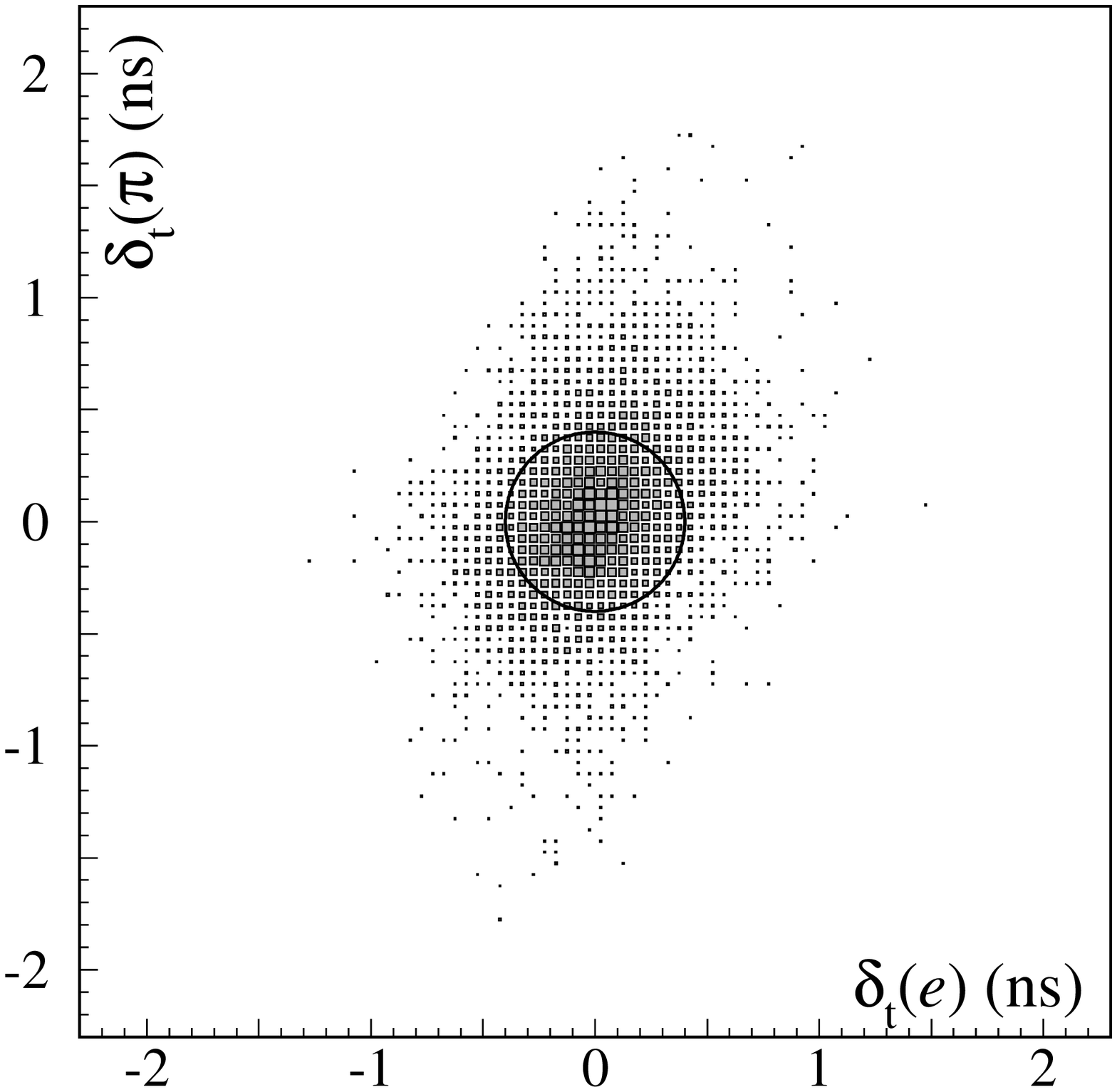, width=0.48\textwidth}
 \caption{Scatter plot of the time differences 
 $\delta_{t}(\Ppi)$ vs $\delta_{t}(\Pe)$
 for $\Ppi$ and $\Pe$ mass assignments for Monte Carlo events, for all \ks\ decays (left)
 and for \DKSeIII\ decays (right). Events within the circles are retained.}
 \label{dtedtp}
\end{figure}

Once particle identification has been performed, 
we reevaluate 
the time differences $\delta_t(m)$, this time using for each track the 
mass assignment known from the cut on $d\delta_{t,e\pi}$ and
subtracting the \tzero\ of the event. For the \tzero\ determination, the bunch crossing producing the event is 
evaluated as the integer part of the ratio $[\delta_t(e)+\delta_t(\pi)]/2T_\mathrm{bunch}$.
We apply another TOF cut by selecting the events within the circle
in the $\delta_t(\Pe)$-$\delta_t(\pi)$ plane,
as shown in \Fig{\ref{dtedtp}} for MC events.
This cut improves the background rejection by a factor of five, while eliminating 8\% of the signal events at this analysis stage.

\begin{figure}
  \center
  \epsfig{file=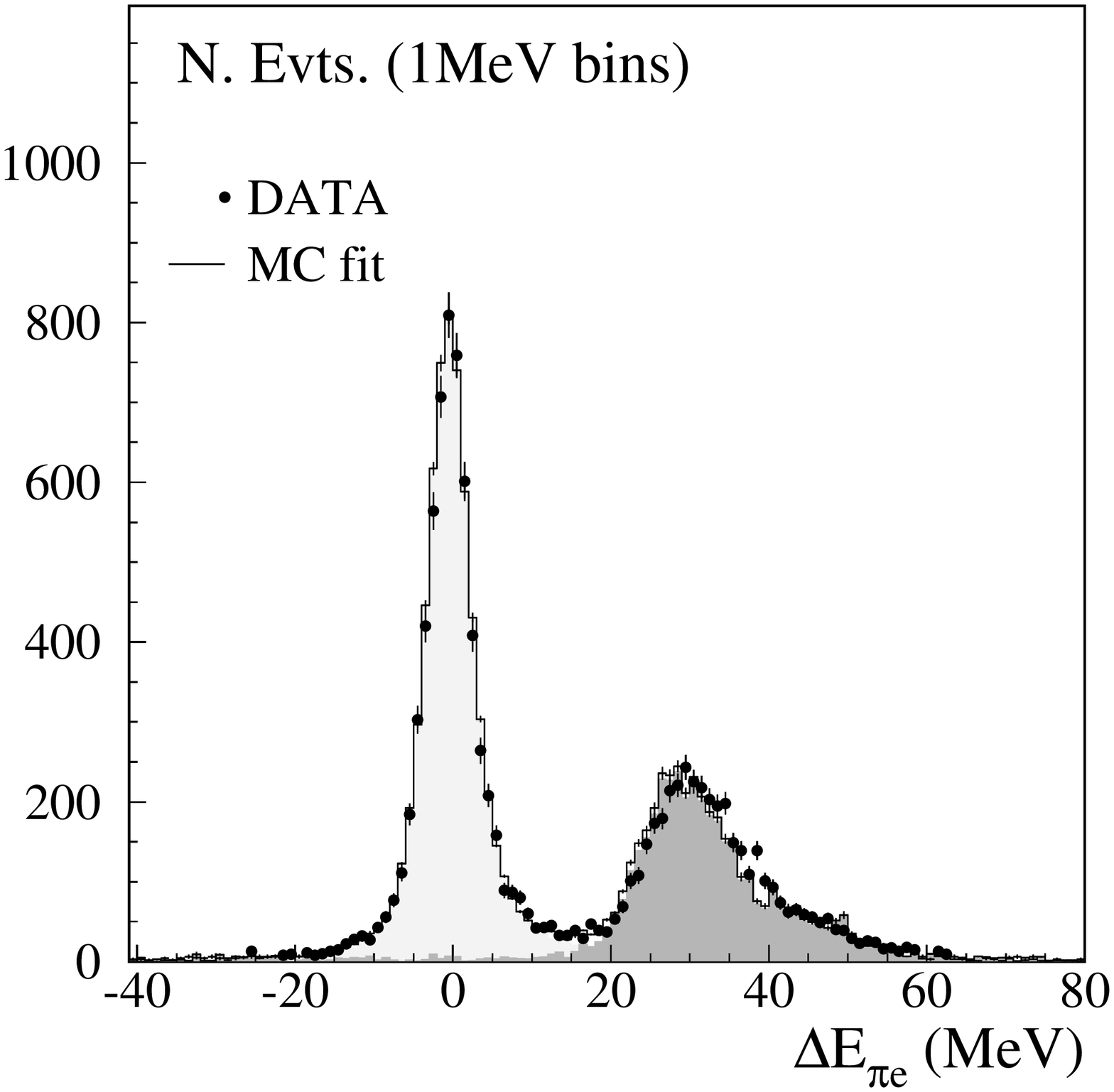, width=0.48\textwidth}
  \epsfig{file=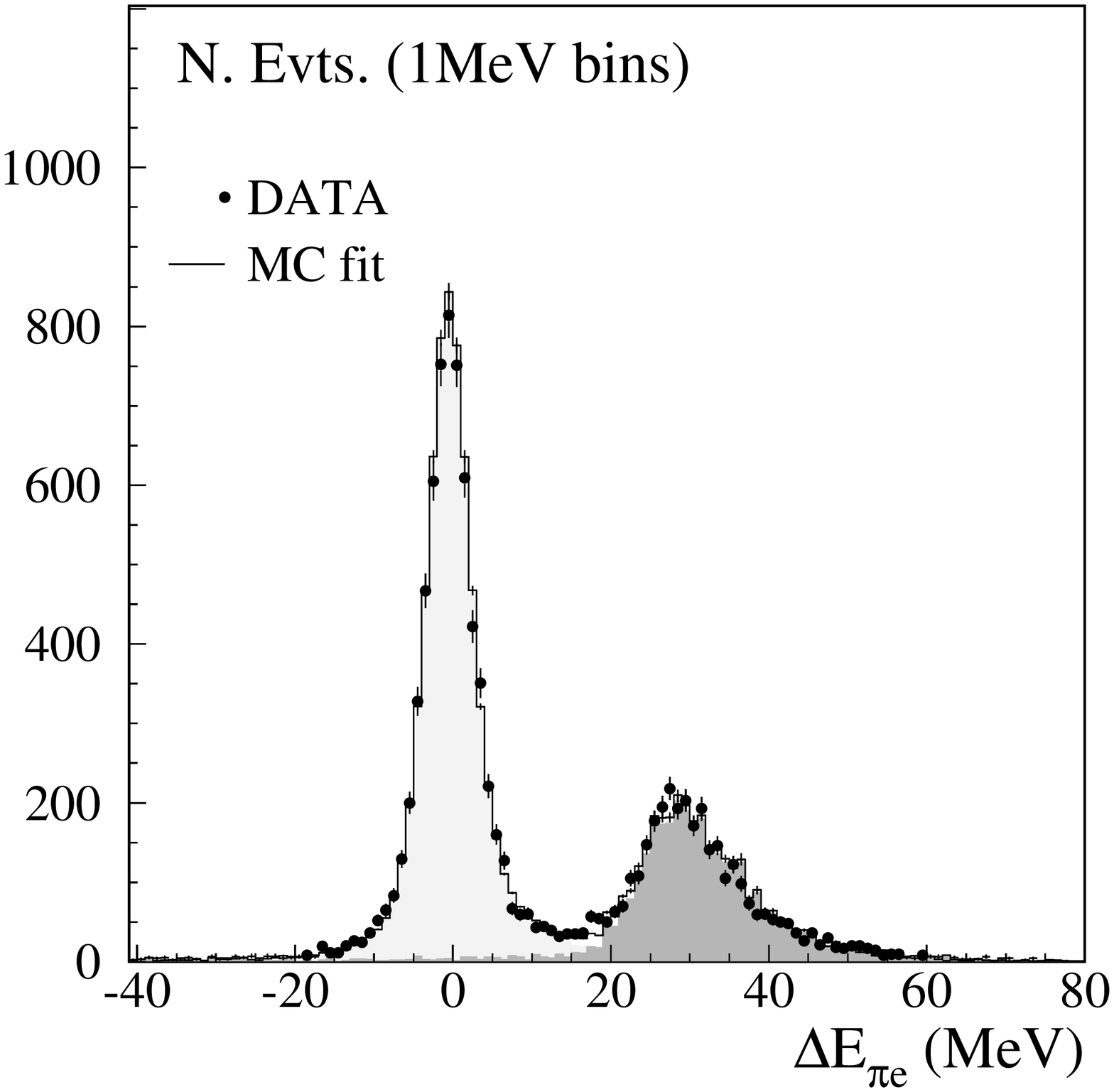, width=0.48\textwidth}
  \caption{$\Emiss-\Pmiss$ spectrum for events selected as $\Ppim\Pep\Pnu$ 
    (left panel) and $\Ppip\Pem\Pnubar$ (right panel). Filled dots represent data
    from the entire data set; the solid line is the result of a fit varying the normalization of 
    MC distributions for signal (light gray) and background (dark gray), which are also shown.}
  \label{fig:Ke3}
\end{figure}

A powerful discriminating variable is the difference between the missing energy and momentum, $\Delta E_{\pi e}\!=\!\Emiss-\Pmiss$,
which is evaluated 
using the \ks\ momentum known from the \kl\ direction. 
For $\pi e \nu$ decays, \Emiss\ and \Pmiss\ are the neutrino energy and momentum, and 
are equal. The distribution of $\Delta E_{\pi e}$ is shown in \Fig{\ref{fig:Ke3}} 
after TOF cuts are imposed for $\Ppim\Pep\Pnu$
(left panel) and for $\Ppip\Pem\Pnubar$  (right panel) candidate events.
A clear peak around zero is evident and corresponds to a clean signal for \DKSeIII.

The residual background is dominated by \DKSpippimandg\ decays. 
Events with $\Delta E_{\pi e}\!>\!10$\MeV\ are mostly due to 
cases in which one pion decays to a muon before entering the tracking volume (``$\pi\mu$'' events), 
in which the track identified as electron by TOF is badly reconstructed  (``$\pi_\mathrm{bad}\pi$'' events), 
or in which the radiated photon has an energy in the \ks\ frame above 7\MeV, thus shifting 
$M_{\pi\pi}$ below 490\MeV\ and 
$\Emiss$ toward positive values (``$\pi\pi\gamma$'' events).
Events with $\Delta E_{\pi e}\!<\!-10$\MeV\ are mostly ``$\pi\mu$'' 
or ``$\pi_\mathrm{bad}\pi$''
events, or are due to cases in which the track identified as the pion by TOF is badly reconstructed
(``$\pi\pi_\mathrm{bad}$'' events).

We discriminate between signal and residual background events by means
of 5 kinematic variables: $\Delta E_{\pi e}$ itself;
the difference $\mathrm{d}_\mathrm{PCA}$  between the impact parameters 
of the two tracks with the IP;
the difference $\Delta E_{\pi\pi}$ 
in the $\pi$-$\pi$ hypothesis;
the squared mass $M^{2}_\mathrm{trk}(\Pe)$ of track 1(2) when it is identified as electron from TOF, 
calculated assuming that $p_S - p_{2(1)}$ is the momentum of an undetected
pion and that $(p_{S}-p_{1}-p_{2})^{2}=0$;
the energy $E^{*}_{\pi(e)}$ 
of the track identified as a $\pi$ ($e$) from TOF, calculated in the \ks\ rest frame using the pion mass hypothesis.

\begin{figure}[ht]
 \center
 \epsfig{file=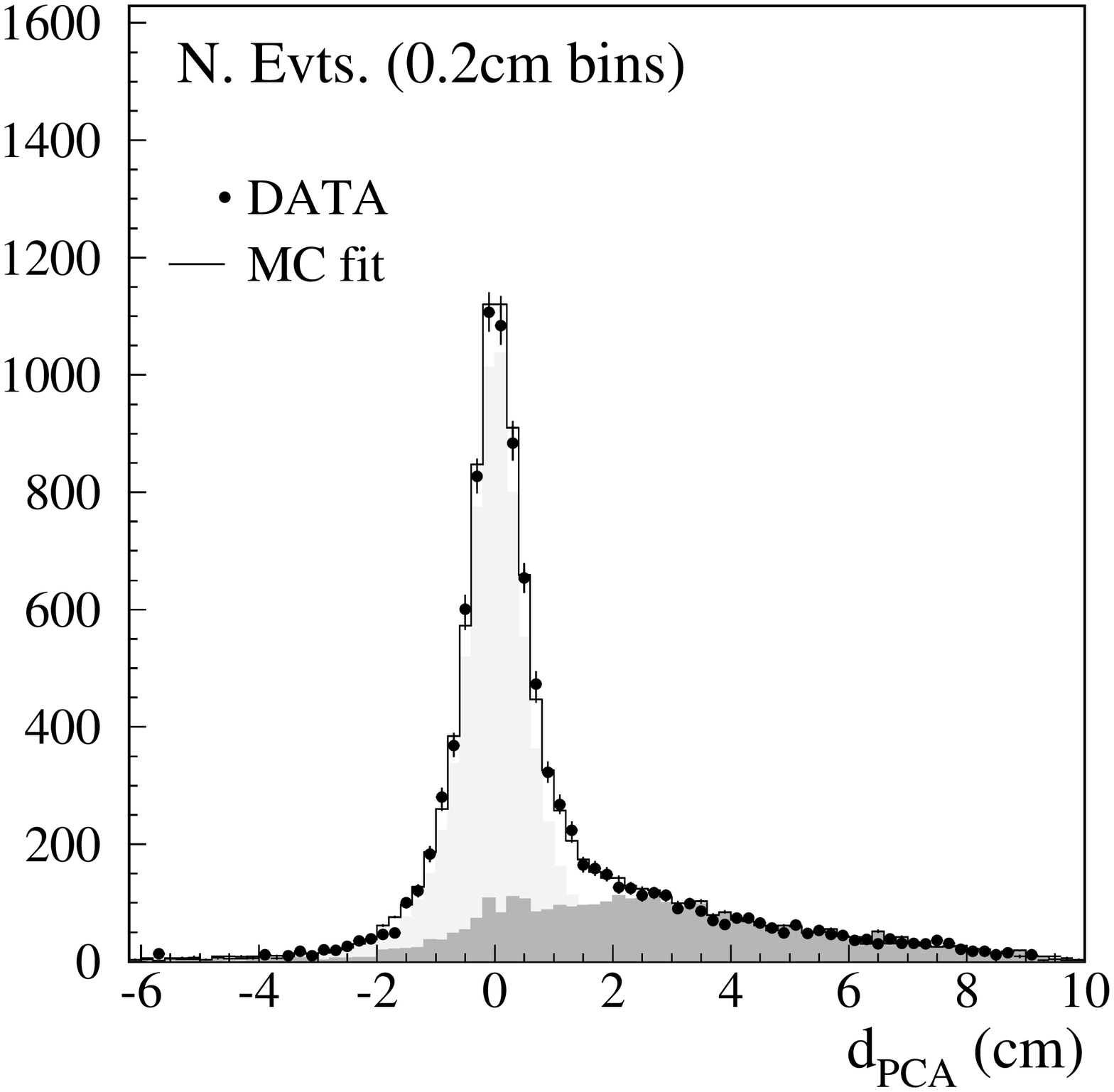, width=0.48\textwidth}
 \epsfig{file=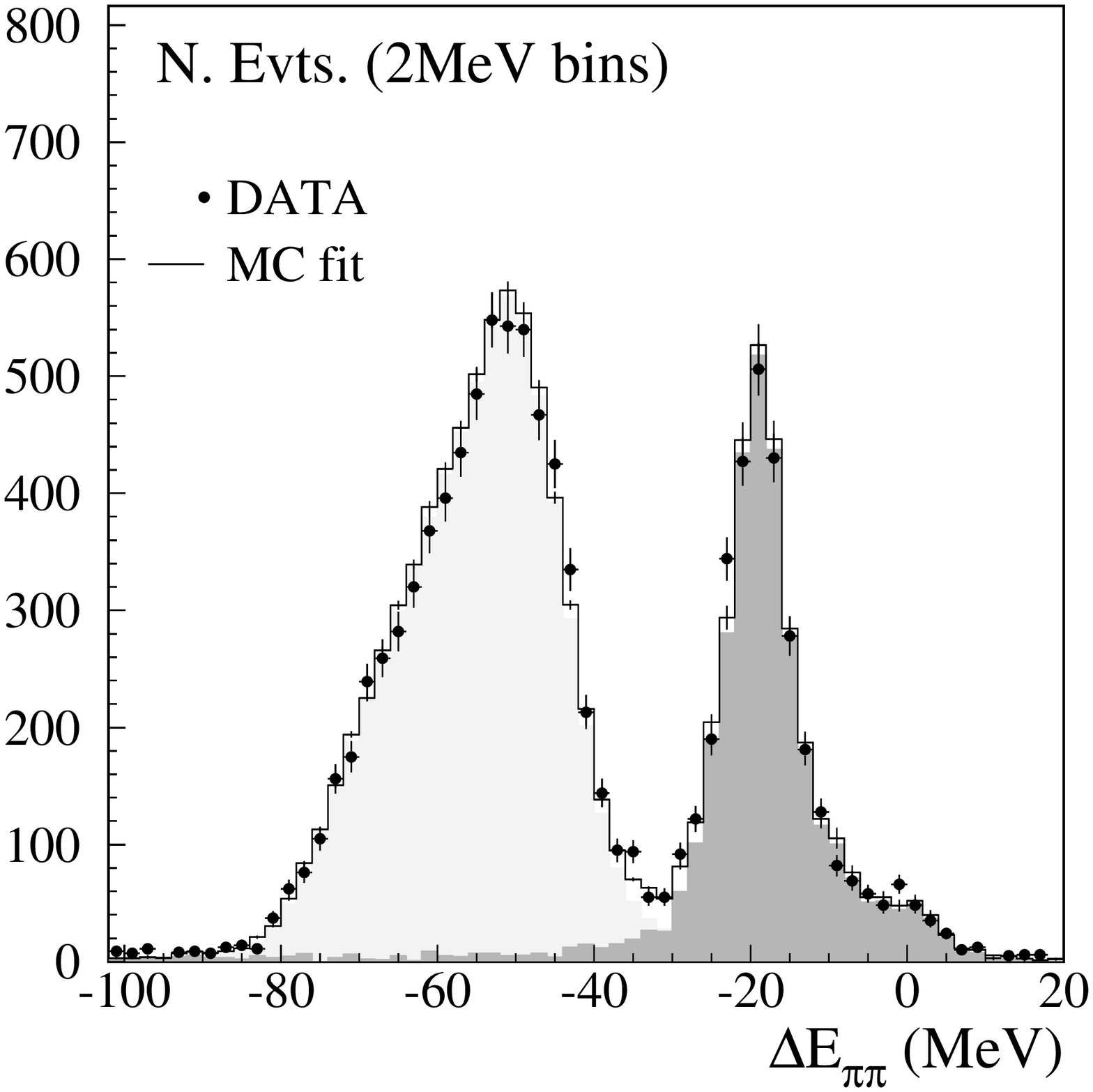, width=0.48\textwidth}
 \epsfig{file=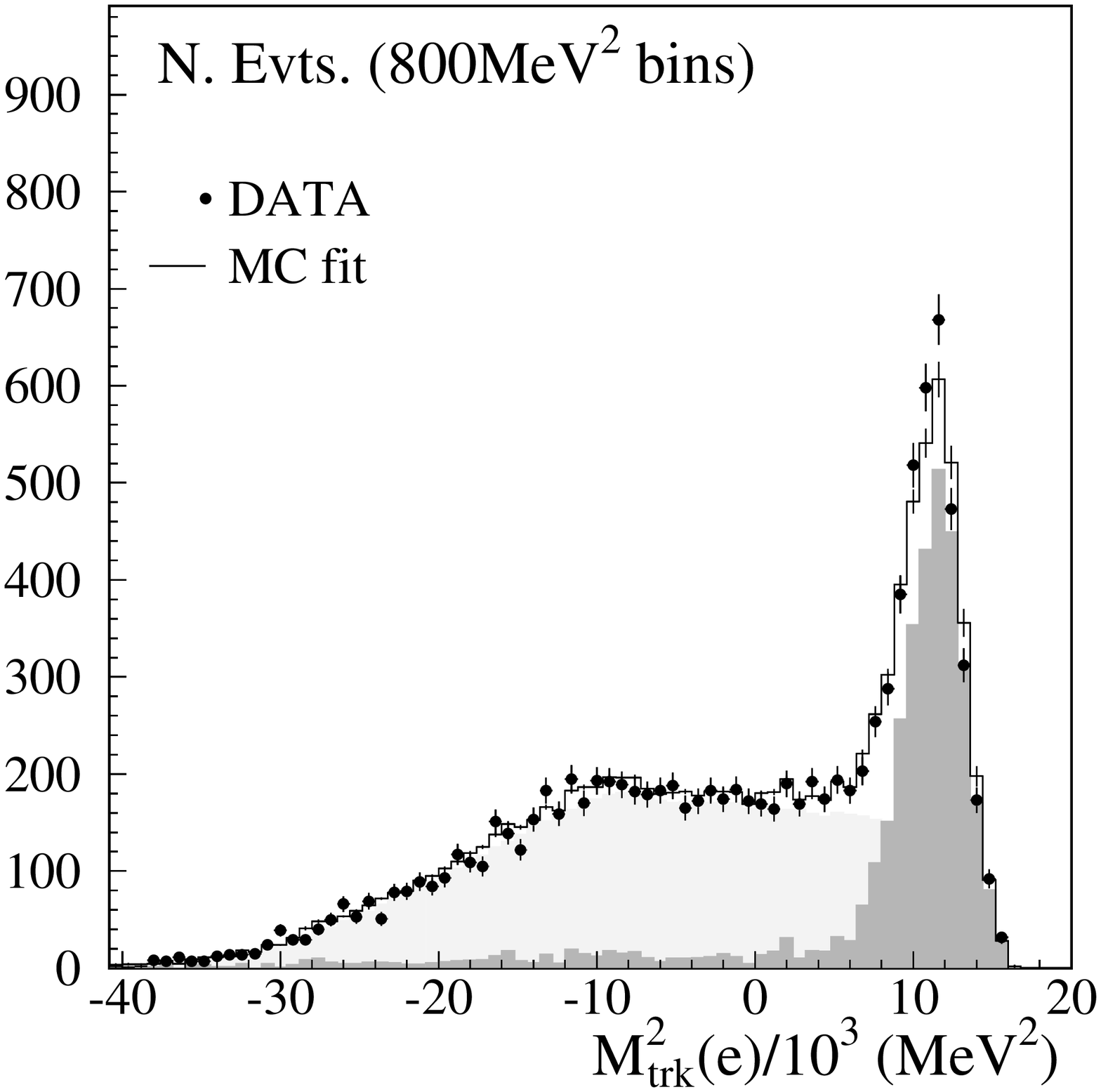, width=0.48\textwidth}
 \epsfig{file=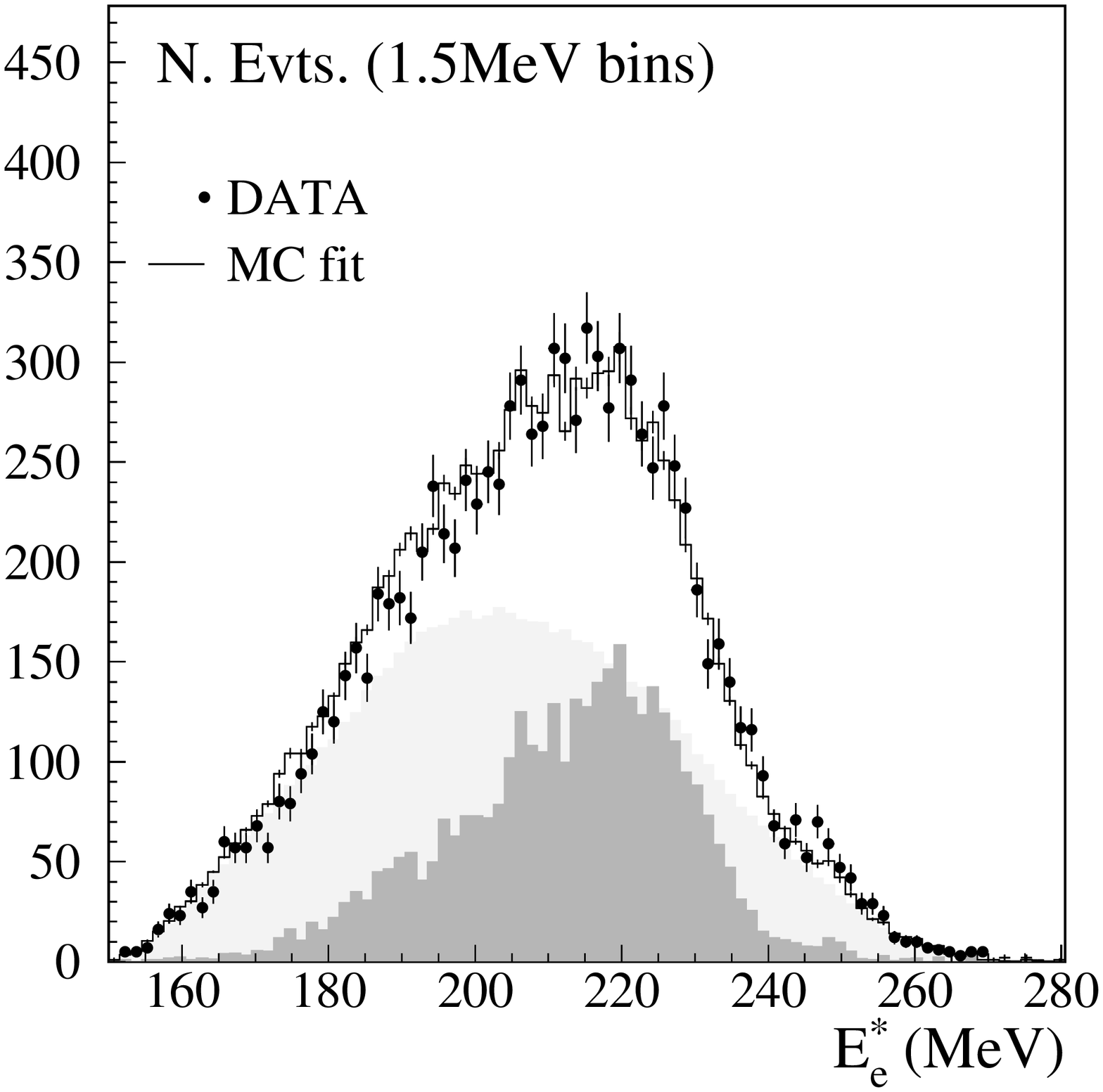, width=0.48\textwidth}
 \caption{Distributions for $\Ppip\Pem\Pnubar$ candidate events, for data (dots) 
and Monte Carlo events (solid line). Top left: $\mathrm{d}_\mathrm{PCA}$ variable, peaking around zero for vertices of good quality 
(signal and $\pi\pi\gamma$ events); top right: $\Delta E_{\pi\pi}$, peaking around zero for $\pi\pi\gamma$ events;
bottom left:  $M^{2}_\mathrm{trk}(\Pe)$, peaking around $m_{\mu}^{2}$ for $\pi\mu$ events; bottom right: $E^{*}_{e}$, peaking around $m_K/2$ for 
$\pi\pi_\mathrm{bad}$ events. In each plot, signal (light gray) and background (dark gray) contributions are also shown.}
 \label{fig:dposmtrk}
\end{figure}

Except for ``$\pi\pi\gamma$'' events, all of the background categories 
are characterized by poor vertex reconstruction quality, leading to
a broad distribution of $\mathrm{d}_\mathrm{PCA}$ as shown in 
\Fig{\ref{fig:dposmtrk}}, top left. 
In contrast, signal and  ``$\pi\pi\gamma$'' events are peaked around zero.
We discriminate ``$\pi\mu$'' events by the $M^{2}_\mathrm{trk}(\Pe)$ variable, 
which peaks around $m_{\mu}^{2}$, and
``$\pi\pi\gamma$'' events by the $\Delta E_{\pi\pi}$ variable, which peaks around zero.
Finally, well reconstructed pion tracks from \DKSpippim\ decays are identified
by the value of $E^{*}$, which peaks around $m_{K}/2$, allowing us to
recognize  $\pi_\mathrm{bad}\pi$ or  $\pi\pi_\mathrm{bad}$ events.

\begin{figure}[htbp]
  \begin{center}
    \rotatebox{270}{\includegraphics[totalheight=10.cm]{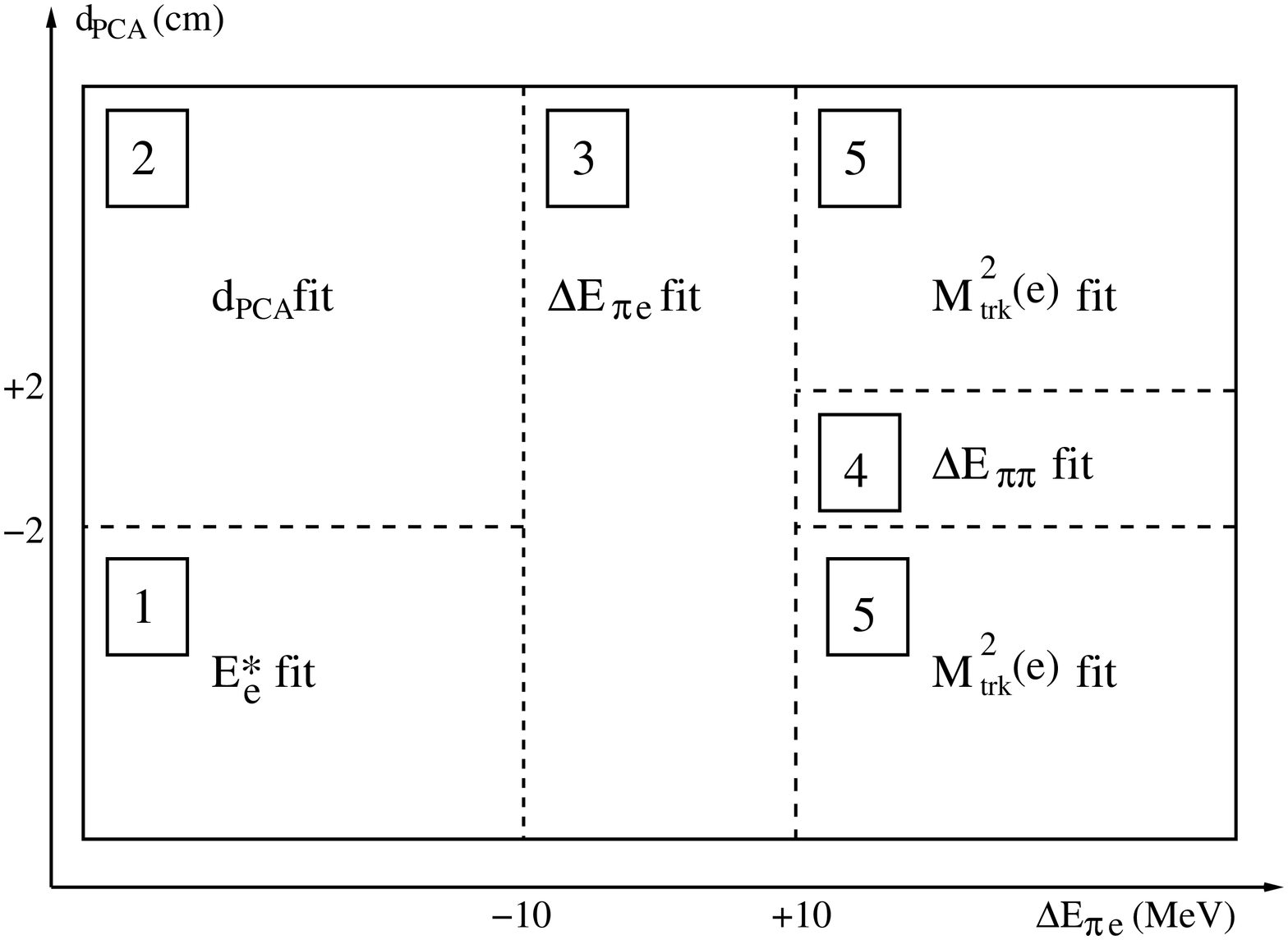}}
    \caption{Definition of the fit regions in the $\Delta E_{\pi e}$-$\mathrm{d}_\mathrm{PCA}$ plane. 
     In each region, the variable used for the fit
     is also specified.}
    \label{fig:planededpos}
  \end{center}
\end{figure}

The number of events due to signal and 
to each of the background categories 
are evaluated through a global binned-likelihood fit using 
the above variables. Each event is assigned to one of five regions in the 
$\Delta E_{\pi e}$-$\mathrm{d}_\mathrm{PCA}$ plane illustrated 
in \Fig{\ref{fig:planededpos}}. For each region, we use for the fit one of the variables defined above.
The choice of the regions and the assignment of the fit variables to each region ensure 
good separation between each component in turn and all the others.
In each region, the data are fit with the sum of the MC distributions in the appropriate variable for
signal events and for 
events from each background source.
The free parameters are the signal and background normalizations. 
For each source, 
the same normalization parameter is used in all fit regions.

The result of the fit is shown as the solid line in the distributions of Figs.~\ref{fig:Ke3} and~\ref{fig:dposmtrk}.
The MC simulations of $\DKSeIII$ and $\DKSpippim$ decays 
include photon radiation in the final state~\cite{radiativearticle}.
If the effect of radiation were
not taken into account, the result for the branching ratio 
would decrease by a few percent.

We perform two independent fits, one for each charge state.
The estimated numbers of signal events are
shown in Tab.~\ref{yields}. The quoted statistical errors include the contributions 
from fluctuations in the signal statistics, 
from the background subtraction, 
and from the finite MC statistics~\cite{Barlow:1993dm}. 

The systematic errors include
the contribution from uncertainty in the shape of the signal distributions.
In particular, we have studied in detail the reliability of the MC in reproducing the distribution of $\Delta E_{\pi e}$. 
We have compared data and MC resolutions obtained from samples of $\DKSpippim$ events tagged by the \kl\ crash, 
both for the momentum of each track and
for the $\Delta E_{\pi\pi}$ variable. From these studies, we have extracted corrections to the MC resolutions for the tracking momentum
and the polar and azimuthal angles for the \kl\ crash.
After these corrections are applied, good agreement is observed for the core of the $\Delta E_{\pi e}$ 
distribution for signal events. In order to study
the reliability of the simulation for the tails, 
an additional variable $x_\mathrm{PID}$ has been used to discriminate signal events.
$x_\mathrm{PID}$ identifies $e^\pm$ tracks on the basis of the spatial distribution of the energy deposition in the EmC, 
and is independent from the track momenta used to obtain the fit variables.
An alternative estimate of the number of signal events in each fit region is obtained from the $x_\mathrm{PID}$ distribution, 
which is reliably reproduced by the MC simulation.
Using this method, a significant difference in the number of signal events 
is observed only in regions 1 and 2, from which a shape correction is defined. This
corresponds to a 0.5\% correction to the final result, which is also taken as the corresponding systematic error.

The cut on the minimum cluster energy required for \kl-crash tagging dramatically affects the fraction of signal events
in the tails of the $\Delta E_{\pi e}$ distribution: the looser the cut, the worse the resolution 
on the \ks\ momentum evaluated from the \kl\ direction.
In order to check the robustness of the signal extraction, we have 
compared results obtained using three values for the minimum \kl\ cluster energy:
125\MeV, 200\MeV, and 300\MeV. The results obtained using independent samples are compatible with each other. 
The intermediate cut gives the minimum total uncertainty.

A detailed description of the fit procedure is given in Ref.~\citen{pennote2}.
\section{Efficiency estimates}
For both \DKSpippim\ (normalization) and \DKSeIII\ (signal) events, 
we estimate the corrections
for the geometrical acceptance and
tagging and selection inefficiencies with the MC simulation.
In order to account for data-MC differences, we weight each event with the ratio of the data and MC tracking efficiencies extracted from
various control samples. We evaluate the probabilities for finding EmC clusters 
from \ks\ daughter particles and for satisfying the trigger conditions by combining data-extracted efficiencies,
parametrized in terms of track momenta, with MC kinematics.
For \DKSeIII\ events, we also use prompt \DKLeIII\ decays 
tagged by \DKSpippim\
as a data control sample for efficiency evaluation.
With this method, we obtain alternative estimates for the trigger and cluster efficiencies and evaluate the corrections
for vertex reconstruction and $\Ppi$-$\Pe$ TOF identification inefficiencies.
These methods are described in detail in Refs.~\citen{pennote2} and~\citen{pennote}.

For \DKSeIII\ decays, the efficiencies are determined separately for each charge 
state and for the 2001 and 2002 data.
We summarize the results for the overall efficiencies, given the tag requirement, in Tab.~\ref{effone}. 
The differences between the efficiencies for the two charge states
arise from the different response of the calorimeter to \Ppip\ and \Ppim,
influencing the cluster, trigger, and TOF efficiencies. 
The uncertainties on the tracking and cluster/trigger efficiencies contribute 
approximately equally to the systematic errors on the overall efficiencies.
A variation in the cluster/trigger efficiencies between 2001 and 2002 
is reflected in the values for the overall efficiencies. 
The corresponding systematic errors have been estimated from the comparison of the 
results obtained using prompt \DKLeIII\ decays and using single-particle weights to 
correct the MC. The difference between the results obtained with the two methods is larger for 
$\pi^-e^+\nu$ events, and for the 2002 data.

\begin{table}
\centering
\renewcommand{\arraystretch}{1.1}    
  \begin{tabular}{ccc}\hline
    $\ks$ decay & \multicolumn{2}{c}{Selection efficiency} \\ 
                     &  Year 2001     & Year 2002 \\ \hline
    $\Ppip\Ppim$ &  
$0.5954 \pm 0.0004_{\mbox{stat}} \pm 0.0010_{\mbox{syst}}$ &
$0.6035 \pm 0.0004_{\mbox{stat}} \pm 0.0010_{\mbox{syst}}$  \\ \hline
    $\Ppim\Pep\Pnu$ & 
$0.2139 \pm 0.0019_{\mbox{stat}}\pm 0.0014_{\mbox{syst}}$ &
$0.2197 \pm 0.0012_{\mbox{stat}}\pm 0.0021_{\mbox{syst}}$ \\ \hline
    $\Ppip\Pem\Pnubar$ & 
$0.2252 \pm 0.0016_{\mbox{stat}} \pm 0.0009_{\mbox{syst}}$ &
$0.2328 \pm 0.0011_{\mbox{stat}} \pm 0.0011_{\mbox{syst}}$  \\ \hline
  \end{tabular}\vglue3mm
\caption{Selection efficiencies for \DKSpippim\ and \DKSeIII\ decays, for the 2001 and 2002 data sets, given the \kl-crash tag.}
  \label{effone}
\end{table}

In principle, the \kl-crash identification efficiency cancels out in the ratio 
of the number of selected  \DKSeIII\ and \DKSpippim\ events.
In practice, since the event \tzero\ is obtained from the \ks\ 
and the \kl\ is recognized by its time of flight,
there is a small dependence of the \kl-crash identification 
efficiency on the \ks\ decay mode.
A correction for this effect is obtained by studying the accuracy of the 
\tzero\ determination in each case~\cite{pennote2,notappoonew,pennote}.
The ratio $R_{\mathrm{tag}}$ of the tagging 
efficiencies for \DKSeIII\ and \DKSpippim\ is found to differ
from unity by $\ab\!1\%$. 
This effect is included in the 
efficiency values shown in Tab.~\ref{effone}.
\section{Results}
For each charge 
state
and for the data set for each year, we obtain the ratio of BR's by normalizing the number 
of signal events to the number of \DKSpippim\ events
and correcting for the overall selection 
efficiencies:
$$
\frac{\gammo{\ks\toP\pi^{\mp}e^{\pm}\nu(\overline{\nu})}}{\gammo{\DKSpippim}}=
\frac{N(\pi^{\mp}e^{\pm}\nu(\overline{\nu}))}{N(\Ppi\Ppi)}\times
\frac{\epsilon_{\mathrm{tot}}^{\Ppi\Ppi}}
{\epsilon_{\mathrm{tot}}^{\pm}}.
$$
The results for the BR's from the data sets for each year are compatible with probabilities greater than 50\%. 
Averaging the results obtained for each data set, we obtain the following results:
\begin{equation}\eqalign{
R_{e+}\equiv\frac{\gammo{\ks\toP\Ppim\Pep\Pnu}}{\gammo{\DKSpippim}} & = \SN{(\VS{\VS{5.099}{0.082_{\rm stat}}}{0.039_{\rm syst}})}{-4}\cr
R_{e-}\equiv\frac{\gammo{\ks\toP\Ppip\Pem\Pnubar}}{\gammo{\DKSpippim}}&=\SN{(\VS{\VS{5.083}{0.073_{\rm stat}}}{0.042_{\rm syst}})}{-4}\cr
R_{e}\equiv\frac{\gammo{\ks\toP\pi e\nu}}{\gammo{\DKSpippim}} &= \SN{(\VS{\VS{10.19}{0.11_{\rm stat}}}{0.07_{\rm syst}})}{-4}\cr}
\label{eq:ratiosbr}
\end{equation}
To obtain the value of the ratio $\BR{\ks\toP\pi e\nu}/\BR{\DKSpippim}$ 
we take into account the correlation between the values measured for
the two charge modes. This correlation arises from uncertainties
on the shapes of the signal distributions in the fit variables that are common to both charge states;
the correlation parameter is 13\%.

The charge asymmetry of Eq.~(\ref{eq:asy}) is given by:
$$
A_{S}\!=\!\frac{
  N(\Ppim\Pep\nu)/\epsilon_{\mathrm{tot}}^{+} -
    N(\Ppip\Pem\bar{\nu}) / \epsilon_{\mathrm{tot}}^{-}} 
     {N(\Ppim\Pep\nu)/\epsilon_{\mathrm{tot}}^{+} +
       N(\Ppip\Pem\bar{\nu})/\epsilon_{\mathrm{tot}}^{-}}.
$$
Combining the results for all data, we obtain:
$$
A_{S}\!=\!\pt(1.5\pm9.6_{\mathrm{stat}}\pm2.9_{\mathrm{syst}}),-3,.
$$
In order to perform a stability test, we have divided the entire
data set into 17 samples and performed the analysis individually for each sample. Values of $\chi^{2}$ corresponding to probabilities
above 50\% are observed for all of the measured quantities~\cite{pennote2}.

The various contributions to the total fractional error on \BR{\DKSeIII} and to the total error on $A_{S}$ 
are listed in Tabs.~\ref{tab:megaerrorbr} and~\ref{tab:megaerrorasy}.
For the measurement of the BR, the uncertainty on the signal count from fit systematics is the dominant contribution to 
the total systematic error.
\begin{table}
\centering
\renewcommand{\arraystretch}{1.1}    
    \begin{tabular}{lcc}\hline
                                             &  \multicolumn{2}{c}{Fractional error ($10^{-3}$)} \\[-1mm] 
                                             &  Statistical  & Systematic    \\ \hline
      Statistics of \DKSeIII                 &  9.1   & \\ 
      Statistics of \DKSpippim               &  0.1   & \\ 
      Preselection efficiency, \DKSeIII      &  1.5   & 2.9    \\ 
      Trigger efficiency, \DKSeIII           &  0.2   & 0.3   \\ 
      TOF efficiency, \DKSeIII               &  2.3   &        \\ 
      Fit systematics, \DKSeIII              &        & 6.2    \\ 
      Preselection efficiency, \DKSpippim    &  0.3   & 1.6    \\ 
      Trigger efficiency, \DKSpippim         &  0.1   & 0.8    \\ 
      Ratio of tagging efficiencies          &  5.0   &        \\ 
      Ratio of cosmic veto inefficiencies    &  1.0   &        \\ \hline
      Total                                  &  10.8  & 7.1    \\ \hline
      {\bf Total fractional error}          & \multicolumn{2}{c}{\bf 12.9} \\ \hline
    \end{tabular}\\[2mm]
  \caption{Contributions to the fractional error on $\BR{\DKSeIII}/\BR{\DKSpippim}$.}
  \label{tab:megaerrorbr}
\end{table} 
\begin{table}
  \begin{center}
\renewcommand{\arraystretch}{1.1}    
    \vglue4mm\begin{tabular}{lcc}\hline
                                             &  \multicolumn{2}{c}{Error ($10^{-3}$)} \\[-1mm]
                                             &  Statistical  & Systematic               \\ \hline
      Statistics of \DKSeIII\                &  9.1   &  \\ 
      Preselection efficiency, \DKSeIII      &  1.5   & 2.9    \\ 
      Trigger efficiency, \DKSeIII           &  0.1   & 0.3    \\ 
      TOF efficiency, \DKSeIII               &  2.3  &     \\
      Fit systematics, \DKSeIII              &       & 0.4 \\ 
      Tagging efficiencies                   &  0.4 & \\ 
      Cosmic veto inefficiencies             &  1.0 &  \\ \hline
      Total                                  &  9.6   & 2.9       \\ \hline
      {\bf Total error}                      & \multicolumn{2}{c}{\bf 10.0} \\ \hline
    \end{tabular}\\[2mm]
  \end{center}
  \caption{Contributions to the absolute error on $A_{S}$.} 
  \label{tab:megaerrorasy}
\end{table} 

For the purposes of measuring the dependence of the form factor $f_{+}(t)/f_{+}(0)$ on the
4-momentum transfer squared $t=(P_{K}-P_{\pi})^{2}$, the elimination of background from 
the sample while preserving statistics is a more important consideration than the 
understanding of the selection
efficiencies at the level required in the branching-ratio analysis. We therefore
use slightly different selection criteria to isolate a clean sample of \DKSeIII\ events:
we require $|\Delta E_{\pi e}|\!<\!10\MeV$ and cut on the quality of the \ks\ vertex.
In order to limit loss of statistics, we loosen the energy requirement on the \kl\ crash to 125\MeV.
We select about 15\,000 signal events, combining the data for the 
two years and charge modes. With this selection, 
the bacgkround contamination 
is reduced to 0.7\%. 
Because of the limited statistics, we only
measure the slope parameter of the form factor in the linear approximation, $f_{+}(t)/f_{+}(0)=1+\lambda_{+}t/m_{\pi}^{2}.$
More precisely, we fit the ratio of data and MC distributions in $t/m_{\pi+}^{2}$ with the function:
\begin{equation}
  \nonumber
  F(t)= A \times\left(\frac{1+\lambda_{+} t/m_{\pi+}^{2}}{1+\lambda_{+,MC} t/m_{\pi+}^{2}}\right)^{2},
\end{equation}
where $A$ and $\lambda_{+}$ are the free parameters of the fit, and 
$\lambda_{+,MC}=0.03$ is the 
value of the slope 
used in the MC generation. Effects from the finite resolution on $t$
are negligible with respect to the statistical error and are ignored.
We find $\lambda_{+} = (33.9\pm4.1)\times10^{-3}$ 
with $\chi^{2}/\mathrm{dof}=12.9/11$, corresponding to a probability 
$P(\chi^{2})\simeq30\%$. This result is in reasonable agreement with the value of $\lambda_{+}$
for semileptonic \kl\ and $K^+$ decays,
$(28.82\pm0.34)\times10^{-3}$,
from the average of results from KTeV~\cite{KTeVff:04}, ISTRA+~\cite{ISTRA+ff:04}, NA48~\cite{NA48ff:04}, and KLOE.~\cite{KLOEff:06}.

\section{Interpretation of the results}
\subsection{Determination of absolute BR's}
In order to evaluate the BR's for the semileptonic modes,
we combine the ratios of BR's measured for each charge [Eq.~(\ref{eq:ratiosbr})] with
the most precise measurement of the ratio
\begin{equation}
\label{eq:rappold}
\R = \frac{\gammo{\DKSpippim}}{\gammo{\DKSpiopio}}=2.236\pm0.015,
\end{equation}
which was also obtained at KLOE~\cite{plb_rappo}.
The only remaining mode with a BR large enough to measurably affect the constraint $\sum_{f}\BR{\ks\!\toP\!f}\!=\!1$ is $K_{\mu3}$; the BR's for 
all other channels sum up to $\ab10^{-5}$. 
Assuming lepton universality, we have 
\begin{equation}
r_{\mu e}=\frac{\Gamma\left(\DKSmuIII\right)}{\Gamma\left(\DKSeIII\right)}=\frac{1+\delta_{K}^{\mu}}{1+\delta_{K}^{e}}\frac{I_{K}^{\mu}}{I_{K}^{e}},
\end{equation}
where $\delta_{K}^{\mu,e}$ are mode-dependent long-distance radiative corrections and $I_{K}^{\mu,e}$ are 
decay phase-space integrals. Using $I_{K}^{\mu}/I_{K}^{e}=0.6622(18)$ 
from KTeV~\cite{Alexopoulos:2004sy} and $(1+\delta_{K}^{\mu})/(1+\delta_{K}^{e})=1.0058(10)$ 
from Ref.~\citen{Andre:2004tk}, we get $r_{\mu e}=0.6660(19)$.
We evaluate the four main BR's of the \ks\ from
\begin{equation}
\label{eq:bratioresults}
\BR{\ks\to i}=\frac{\gammo{\ks\to i}/\gammo{\DKSpippim}}{1+1/\R+(R_{e+}+R_{e-})(1+r_{\mu e})},
\end{equation}
where $i\!=\!\pi^{+}\pi^{-},\,\pi^{0}\pi^{0},\,\Ppim\Pep\Pnu,\,\Ppip\Pem\Pnubar$.
We find: 
\begin{equation}\label{res:brs}\eqalign{
\BR{\DKSpippim}&=(69.02\pm 0.14 )\x10^{-2}\cr
\BR{\DKSpiopio}&=(30.87\pm 0.14 )\x10^{-2}\cr
\BR{\DKSeIIIeppm}&=(3.519\pm 0.063)\x10^{-4}\cr
\BR{\DKSeIIIempp}&=(3.508\pm 0.058)\x10^{-4}\cr}
\end{equation}
The correlation matrix 
$\langle\delta_{i}\delta_{j}\rangle/\sqrt{\langle\delta_{i}^{2}\rangle\langle\delta_{j}^{2}\rangle}$ 
is:
\begin{equation}
  \begin{array}{cc}
    &
    \begin{array}{cccc}
      \pi^{+}\pi^{-}\kern3mm & 
      \pi^{0}\pi^{0}\kern3mm & 
      \kern1mm\pi^{-}e^{+}\nu\kern2mm & 
      \kern1mm\pi^{+}e^{-}\overline{\nu}\kern1mm\\
    \end{array}\\
    \begin{array}{c}
      \pi^{+}\pi^{-} \\ 
      \pi^{0}\pi^{0} \\
      \pi^{-}e^{+}\nu \\ 
      \pi^{+}e^{-}\overline{\nu}\\
    \end{array}
    &
    \left(
    \begin{array}{cccc}
   1      &
  -0.9999 &
   0.1106 &
   0.1196 \\
  -0.9999 &
   1	  &
  -0.1187 &
  -0.1272 \\
   0.1106 &
  -0.1187 &
   1	  &
   0.1445 \\
   0.1196 &
  -0.1272 &
   0.1445 &
   1      \\
    \end{array}
    \right)
  \end{array}
\end{equation}
The contribution from the error on $r_{\mu e}$ is included in the systematic errors. 
Taking 
correlations into account, we have:
\begin{equation}
\label{res:brke3}
  \BR{\DKSeIII}  = \left(7.028\pm0.092\right) \times 10^{-4}.
\end{equation}

\subsection{Test of the $\Delta S=\Delta Q$ rule}
From the total BR we test the validity of the $\Delta S=\Delta Q$ rule in $CPT$-conserving 
transitions [Eq.~(\ref{rex})]. 
We use the following values for the \ks\ and \kl\ lifetimes:
$\tau_{S} = 0.08958(6)$\ns\ from  the PDG~\cite{PDBook} and 
$\tau_{L} = 50.84(23)$\ns\
from 
recent measurements from KLOE~\cite{Ambrosino:2005ec,Ambrosino:2005vx}.
For \BR{\DKLeIII}, we use the value from 
KLOE, 0.4007(15)~\cite{Ambrosino:2005ec}.
We obtain:
\begin{equation}
\label{res:rexplus}
\mathrm{Re}\,x_{+} = \left(-1.2\pm3.6\right)\times 10^{-3}.
\end{equation}
The error on this value represents an improvement by almost a factor of two with respect to the 
most precise previous measurement,
that from the CPLEAR experiment~\cite{CPLEAR_rex:98}.

\subsection{Test of the $CPT$ symmetry}

From the sum and difference of the \kl\ and \ks\ charge asymmetries one can test for possible 
violations of the $CPT$ symmetry, 
either in the decay amplitudes 
or in the mass matrix [Eqs.~(\ref{eq:rexm}) and~(\ref{eq:rey})].
Using $A_L = (3.34\pm0.07)\times10^{-3}$~\cite{PDBook}, we obtain
from Eq.~(\ref{eq:rexm})
\begin{equation}
\mathrm{Re}\,x_{-}+\mathrm{Re}\,\delta=\left(-0.5\pm2.5\right) \times 10^{-3}.
\end{equation}
Current knowledge of these two parameters is dominated by results from CPLEAR \cite{CPLEAR_redelta:98}: the error on $\mathrm{Re}\,\delta$ is 
$3\times10^{-4}$ and that on $\mathrm{Re}\,x_{-}$ is $10^{-2}$. Using   
$\mathrm{Re}\,\delta=(3.0\pm3.3_{\rm stat}\pm0.6_{\rm syst})\times10^{-4}$ from CPLEAR, we obtain:
\begin{equation}
\mathrm{Re}\,x_{-}=\left(-0.8\pm2.5\right) \times 10^{-3},
\end{equation}
thus improving on the error of $\mathrm{Re}\,x_{-}$ by a factor of five.

From Eq.~(\ref{eq:rey}) we obtain
\begin{equation}
\label{eq:reymwnoree}
  \mathrm{Re}\,\epsilon-\mathrm{Re}\,y=\left(1.2\pm2.5\right) \times 10^{-3}.
\end{equation}
We calculate $\mathrm{Re}\,\epsilon$ using values from Ref.~\citen{PDBook} that are obtained without assuming
$CPT$ conservation: 
$\mathrm{Re}\,\epsilon=|\epsilon|\times\cos{\phi_{\epsilon}}=(1.62\pm0.04)\times10^{-3}$.
Subtracting this value from Eq.~(\ref{eq:reymwnoree}), we find
\begin{equation}
  \mathrm{Re}\,y=\left(0.4\pm2.5\right) \times 10^{-3},
\end{equation}
which has precision comparable to that ($3\!\times\!10^{-3}$) obtained from the unitarity relation by CPLEAR~\cite{CPLEAR:bell}.
\subsection{Determination of \Vus}
The value of \Vus\ can be extracted from the measurement of \BR{\ks\toP\pi e\nu}
and from the \ks\ lifetime, $\tau_{S}$:
\begin{equation}
\label{eq:vuss}
\Vus\ \times f_{+}^{K^0 \pi^-} (0) = \left[ \frac{128\,\pi^3\, \BR{\ks\toP\pi e\nu}}
{\tau_{S}\,G_{\mu}^2\, M_{K}^5\,  S_{\mathrm{ew}}\,I_{K}(\lambda_+, 0) } \right]^{1/2}
\frac{1}{1 + \delta^{K}_{{\mathrm{em}}}},
\end{equation}
where $f_{+}^{K^0 \pi^-}(0)$ is the vector form factor at zero momentum
transfer and $I_{K}(\lambda_+, 0)$ is the result of the phase
space integration after factoring out $f_{+}^{K^0 \pi^-}(0)$; both quantities are
evaluated in absence of radiative corrections.
The radiative corrections~\cite{Cirigliano:2004pv,Andre:2004tk} 
for the form factor and the phase-space integral
are included via the parameter 
$\delta^{K}_{{\mathrm{em}}}=(0.55\pm0.10)\!\times\!10^{-2}$~\cite{Cirigliano:2004pv}.
The short-distance electroweak corrections are included in the
parameter $S_{\mathrm{ew}}=1.0232$~\cite{Marciano:1993sh}.

The pole parametrization of the vector form factor is $f_+(t)=f_+(0)[M_V^2/(M_V^2-t)]$.
Expanding this expression to second order gives $\lambda_+''=2\lambda_+'^2$, where 
$\lambda_+'$ and $\lambda_+''$ are the linear and quadratic slopes,
$$
f_{+}(t)=f_{+}(0)\left[1+\lambda_{+}^{\prime}\frac{t}{m^{2}_{\pi^{+}}}+\frac{\lambda^{\prime\prime}_{+}}{2}\frac{t^{2}}{m^{4}_{\pi^{+}}}\right].
$$
We evaluate the phase space integral from the value of $M_V$ from KLOE, $M_V\!=\!870.0\pm9.2$\MeV~\cite{KLOEff:06}, and
get $I_K\!=\!0.10320\pm0.00020$.
The pole fit result is less affected by the strong correlation between the linear and quadratic slopes and
provides better consistency among the values of $I_K$ from different experiments 
(KLOE~\cite{KLOEff:06}, KTeV~\cite{KTeVff:04}, ISTRA+~\cite{ISTRA+ff:04,ISTRA+mu3:04}, and NA48~\cite{NA48ff:04})
than is obtained using the results for $\lambda_+'$ and $\lambda_+''$.

We obtain:
\begin{equation}
\label{res:vusfplus}
f_{+}^{K^0 \pi^-}(0)\times \Vus\ = 0.2150\pm 0.0014
\end{equation}
Using $f_{+}^{K^0 \pi^-}(0)=0.961\pm0.008$ 
from Ref.~\citen{Leutwyler:1984je} (this value is in agreement with a recent lattice calculation~\cite{Becirevic:04}), we get 
\begin{equation}
\label{res:vus}
\Vus\!=\!0.2238\pm0.0024.
\end{equation}
To perform a test of first-row CKM unitarity, we define:
\[
\Delta = 1-\Vud^{2}-\Vus^{2}-\Vub^{2}
\]
Using $\Vud\!=\!0.97377\pm0.00027$ from Ref.~\citen{Marciano:2005ec} and including the small contribution of \Vub~\cite{PDBook}, we obtain
\begin{equation}
\label{res:Delta}
\Delta=\left(1.7\pm1.2\right)\times10^{-3},
\end{equation}
which is compatible (1.4~$\sigma$) with zero.

A new measurement of \R\ has recently been made at KLOE, with a 
precision improved by more than a factor of two with respect to
Eq.~\ref{eq:rappold}. Combining the new and old KLOE measurements, we obtain $\R\!=2.2549\pm0.0054$.
The results presented in Eqs.~\ref{res:brs},~\ref{res:brke3},~\ref{res:rexplus},~\ref{res:vusfplus},~\ref{res:vus}, and~\ref{res:Delta} 
depend only slightly on the value used for \R. The results obtained using the updated value of \R\ are listed below:
  \begin{equation}
\eqalign{
\BR{\DKSpippim}&=(69.196\pm0.051)\x10^{-2} \cr 
\BR{\DKSpiopio}&=(30.687\pm0.051)\x10^{-2} \cr 
\BR{\DKSeIIIeppm}&=(3.528\pm0.062)\x10^{-4}\cr 
\BR{\DKSeIIIempp}&=(3.517\pm0.058)\x10^{-4}\cr 
\BR{\DKSeIII}  &=(7.046\pm0.091) \x 10^{-4}\cr}
\end{equation}
From the last of these, the following quantities follow:
  \begin{equation}
\eqalign{
\mathrm{Re}\,x_{+}&=(-0.5\pm3.6)\x10^{-3}\cr 
f_{+}^{K^0 \pi^-}(0)\times \Vus\ &=  0.2153\pm0.0014\cr
\Vus &=0.2240\pm0.0024\cr
\Delta &=(1.6\pm1.2)\x10^{-3}\cr}
\end{equation}
The differences between these quantities when calculated using the old and new values of \R\ 
are well within the stated uncertainties.

\section*{Acknowledgements}
We thank the \Dafne\ team for their efforts in maintaining low-background 
running conditions and their collaboration during all
data taking. We also thank F. Fortugno for his efforts in ensuring
good operations of the KLOE computing facilities 
and F. Mescia and G. Isidori for their help. This work was
supported in part by DOE grant DE-FG-02-97ER41027; by 
EURODAPHNE, contract FMRX-CT98-0169; by the German Federal Ministry of Education and Research (BMBF) contract 06-KA-957; 
by Graduiertenkolleg 'H.E. Phys. and Part. Astrophys.' of 
Deutsche Forschungsgemeinschaft, Contract No. GK 742;
by INTAS, contracts 96-624, 99-37.
%
\bibliographystyle{elsart-num}
\bibliography{paper}
\end{document}